\documentclass[conference]{IEEEtran}
\usepackage[T1]{fontenc}

\usepackage{ieeefig}
\usepackage{amssymb}
\usepackage{multirow}
\usepackage{bigstrut}
\usepackage{array}
\usepackage{float}

\usepackage{epsfig}

\usepackage{algorithmic} 
\usepackage{algorithm} 

\usepackage[cmex10]{amsmath}
\ifCLASSINFOpdf 
   \usepackage{graphicx}
\else 
\fi   

\begin{document}
%
%
  \title{Optimizing Channel Access for Event-Driven Wireless Sensor Networks:
  Analysis and Enhancements}


\author{\IEEEauthorblockN{Rajeev K. Shakya,
      Yatindra Nath Singh, and Nishchal K. Verma}
\IEEEauthorblockA{Department of Electrical Engineering, Indian Institute of
Technology Kanpur\\
Email: \{rajeevs,ynsingh,nishchal\}@iitk.ac.in}}


\maketitle

\begin{abstract}
We study the problem of medium access control in domain of event-driven wireless
sensor networks (WSNs)~\cite{ref15}. In this kind of WSN, sensor nodes send data
to sink node only when an event occurs in the monitoring area. The nodes in this
kind of WSNs encounter correlated traffic as a subset of nodes start sending
data by sensing a common event simultaneously. We wish to rethink of medium
access control (MAC) for this type of traffic characteristics. For WSNs,
many existing MAC protocols utilize the basic CSMA/CA strategy such as IEEE
802.11 Binary Exponential Backoff (BEB) algorithm to handle the collisions among
packets when more than one node need to access the channel. We show that this
BEB algorithm does not work well without incurring access delay or performance
degradation due to increased number of collisions and retransmissions when nodes
encounter correlated traffic. Based on above observations in mind, We present a
Adaptive Random Backoff (ARB) algorithm that is capable of mitigating the impact
of correlated traffic and capable of minimizing the chance of
collisions. ARB is based on minor modifications of BEB. We show using numerical analysis that
our proposals improve the channel access in terms of latency, throughput, and frame dropping probability as compared with IEEE 802.11 DCF. Simulations using NS-2 network simulator are
conducted to validate the analytical results.

%


\end{abstract}
\begin{keywords}
CSMA/CA, IEEE 802.11, Backoff algorithm, performance analysis, Wireless Sensor
Network. 
\end{keywords}

%
\IEEEpeerreviewmaketitle

\section{Introduction}
The recent developments in wireless sensor networks have enabled low cost, low
power sensor nodes which are capable of sensing, computing and transmitting
sensory data in environments such as surveillance fields, smart homes, offices
and intelligent transportation systems. These sensor nodes cooperatively monitor
physical and environmental conditions such as temperature, sound, vibration, and
pressure etc. WSNs are also used for critical applications in highly dynamic and
hostile environments. Therefore, they must adapt to failure of nodes and loss of
connectivity. 

The sensor networks are emerging area of mobile ad-hoc network that presents
novel networking issues because of their different application requirements,
limited resources capabilities and functionalities, small packet size, and
dynamic multi-hop technologies. The Medium Access Control (MAC) is an important
mechanism to share wireless communication channel by resolving the issues of
collisions thus leading to successful transmission of packets. Many MAC
protocols have been designed for traditional wireless networks~\cite{ref4}.
While these protocols are well suited for traditional wireless networks, they
are not adequate for wireless sensor networks due to different nature of traffic
characteristics and application behavior. The primary function of a sensor
network is to sample sensory information from its vicinity such as temperature,
light, and send this data to the base station node. The base station node mostly
forwards all the data wire-line or an independent wireless network to control
center. The sensor nodes in network operate as a collective structure which
makes this network different than traditional ad-hoc networks. Due to dependency
on sensor for generating data, the traffic characteristics are variable and
correlated. Mostly they change little over long period of time and on other hand
can be very intense for short period of time. This correlated traffic is
characteristics of densely deployed wireless sensor network applications. For
example, in room monitoring application where a fire in a room of a building
triggers a number of nodes attached with temperature sensors to begin reporting
a common event. These all nodes, simultaneously become active and transmit
packets. This behavior causes the nodes to operate in spatially-correlated
contention where multiple nodes of same neighborhood sense a common event at
same time~\cite{ref14,ref15}.

We wish to rethink of CSMA based MAC design for this types of traffic
characteristics in mind especially for event-driven based sensor network
applications. Tay, Jamieson, and Balakrishnan~\cite{ref4} have defined
characteristics for event-driven sensor network applications as follows.
\begin{itemize}
 \item An event-driven based sensor network encounters spatially-correlated
contention, once an event occurs into a particular region/zone of monitored
area. The multiple nodes of same neighborhood sense it and send data to report
it to the base station node. As a result, a synchronized burst of transmissions
happens.
\item In many applications, all the packets need not be treated as equally
important. It is enough that some nodes out of all, are successful in
transmitting the data.
\item The number of nodes getting activated by an event in a particular region
changes with time. For example, when a target enters into a sensor field, the
number of active sensing nodes could become large very quickly. 
\end{itemize}
In these types of traffic patterns, channel access delay and system throughput
are the performance limiting factor when number of contending nodes are large
that are activated by means of occurring an event into a specific zone of sensor
field. Our goal is to minimize the channel access delay and improve the system
throughput in event-driven based sensor network applications.

These characteristics pose a challenge in design of MAC protocol for
event-driven based wireless sensor networks. Tay, Jamieson, and
Balakrishnan~\cite{ref14} have also proposed a solution to this problem using
fixed contention window size of 32 slots with non-uniform probability
distribution. This is contention-based protocol, (named as SIFT) to allow first
R winners of contention to send the reports to base station when N nodes sense
an event and contend to transmit simultaneously. It has been
designed for sensor network applications when spatially-correlated contention
occurs by sensory measurements of an event that triggers many sensor nodes in
event-driven scenario, where multiple sensor nodes detect an event to send data
to the sink node through multiple-hop communication. The basic idea in SIFT
protocol is to randomly selecting contention
slots using non-uniform probability distribution within fixed sized contention
window rather than using variable contention window size as used in many
traditional MAC protocols. Simulation results show that SIFT outperforms 802.11
DCF in term of report latency by a factor of 7~\cite{ref14}. One drawback of
SIFT is that it degrades in performance when number of contending nodes exceeds
512. In other words, as the number of nodes reporting an event become larger
than 512, slots get picked by multiple nodes, resulting in collisions. Authors
argued that traditional carrier-sense multiple access (CSMA) strategies like
IEEE 802.11 DCF is not suitable in this problem domain, and the larger value of
contention window size is not always necessary for sensor network applications
since traffic is mostly very low and becomes intense only when any activity is
observed in the environment.

To mitigate the impact of spatially-correlated contention among the active nodes
on the network performance, particularly observed into event-driven based sensor
applications~\cite{ref14}, we proposed a new adaptive and predictable algorithm,
called adaptive random backoff (ARB), that is based on minor modifications of
the IEEE 802.11 binary exponential backoff (BEB). The motivation of ARB is the
enhancement of BEB performance for correlated traffic generated from multiple
nodes of same neighborhood when these nodes sense an event in a
event-driven scenario. The rest of paper is organized as follows: Section II
illustrates the behavior of IEEE 802.11 DCF. Section III presents the overview
of related work. Section IV describes the details of our proposed algorithm
ARB. Section V evaluates the efficiency of ARB through simulation experiments.
Finally, conclusions and future work are presented in section VI.

\section{IEEE 802.11 $-$ what goes wrong}
In this section we present analysis of IEEE 802.11 DCF method using analytical
model given by~\cite{ref2}. We will show that IEEE 802.11 DCF can cause latency
problem and degrade the performance for event-driven WSN.

The IEEE 802.11 standard~\cite{ref1} is designed to support communication among
nodes in ad-hoc peer-to-peer configuration as well as designation access point
in infrastructure mode of operation. Traditionally, nodes in wireless adhoc
networks, communicate with each other as peers using multi-hop. While in WSNs,
irrespective of multi-hop or single-hop, all the nodes will usually communicate
with a base station node. In the 802.11 protocol, two mechanisms have been
defined for medium access. The fundamental mechanism called, the distributed
co-ordination function (DCF), is a random access scheme based on the Carrier
sense multiple access with collision avoidance (CSMA/CA) protocol. Another
mechanism is defined as an optional point coordination function (PCF) which is a
centralized MAC protocol for supporting collision-free, time bounded services.
 \begin{figure}[!b]
 \centering
 \includegraphics[scale=0.50]{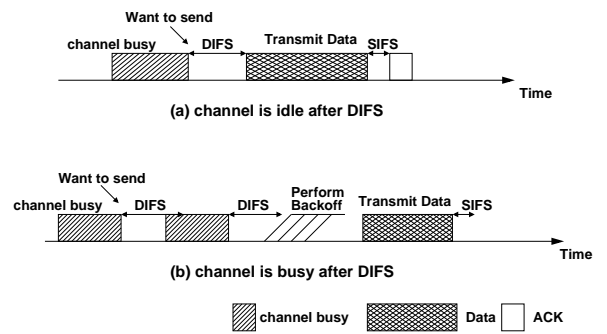}
  \caption{CSMA/CA mechanism illustration}
  \label{fig:BEB}
 \end{figure}
DCF describes two techniques for packet transmission. The default scheme is a
two way hand shaking mechanism called the basic access scheme. Another scheme
called the Request to send/Clear to send (RTS/CTS) is a four way hand shaking
mechanism defined by the standard. It has been specially formulated to encounter
the hidden node/exposed node problem. In basic access scheme, a node, with a
packet to transmit, monitors the channel activity. If the channel is free for a
period of time equal to Distributed Inter Frame Space (DIFS), the node performs
transmission as shown in Fig.\ref{fig:BEB}(a). If the channel is sensed busy,
either immediately, or during the DIFS the backoff procedure is activated. The
node again continues to monitor the channel till it is sensed idle for a DIFS
period of time. At this point, the node generates a random number between 0 and
CW (contention window) and use it to initialize the counter to perform backoff
and starts the countdown procedure. This counter is decremented by one for every
idle time slot sensed by node. When it reaches to zero, node transmits the
packet
immediately as shown in Fig.\ref{fig:BEB}(b). Also a node needs to wait for a
random back-off interval between two of its consecutive successful packet
transmissions by the same node. Initially (i.e. backoff stage 0), backoff is
chosen in between zero to minimum value of contention window ($W_{min}$). After
each unsuccessful transmission due to collisions, backoff stage is incremented
by one and contention window (CW) doubles its value till it reaches to maximum
value of CW ($W_{max}$).According to IEEE 802.11 standard, size of contention
window
will not change when retransmission limit reaches a threshold value, we call it
backoff order, denoted by m. When transmission is successful, the backoff stage
will start with initial value of 0 and initial CW will reset to $W_{min}$,
regardless of network traffic conditions.
The DCF method tends to work well when number of contending nodes is few. As
discussed in previous section, it does not efficiently handle the
spatially-correlated contention occurring in event-driven WSN. To better
understand the performance of DCF, we reproduce the numerical results using
analytical model proposed in~\cite{ref2} and observe the behavior of protocol
under full load conditions.

We observe that, the conditional collision probability increases with the
increase in the number of nodes which causes a reduction in the system
throughput. On the other hand for a single node throughput falls because of
minimum contention window size, which introduces unnecessary idle slots. Based
on observation using analytical model, followings are some of the
drawbacks of the basic access mechanism of IEEE 802.11 DCF:
%
number
fixed

\begin{itemize}
\item Although, DCF tries to avoid frame collision by using the binary
exponential backoff (BEB) algorithm, still it can not avoid the inevitable
retransmission of frames when the number of competing nodes is fairly large.

\item As the waiting time of a node grows exponentially with the number of frame
retransmissions, it becomes mandatory for DCF to bound the number of
retransmissions to a smaller number. 

\item Intuitively we can say that, as contention window size increases, the
average backoff timer value chosen before every transmission attempt, increases.
Thus, on an average, a node will be able to transmit the packet only after a
longer period of time.

\item If the packet at the head of the buffer gets backlogged for a longer
period of time due to consecutive failures in retransmitting the packet, the
remaining packets in the buffer will also have to suffer a larger delay. Many a
times, due to this phenomena, the buffer becomes full with packets and then the
future incoming packets will be dropped till the time the severely backlogged
packet at the head of the queue gets successfully transmitted. 
\end{itemize}
In~\cite{ref2}, an approximate solution relating the transmission probability $
\tau $ with the number of nodes $ \emph N$  is derived, which maximizes the
system
throughput. We can say that $ \tau$ depends only on the system parameters
backoff
order $ \emph m$, contention window CW, and the network size $ \emph N$. Hence,
the way to
achieve optimal performance is to employ adaptive techniques to tune the values
of $ \emph m$ and CW (and consequently $ \tau$) on the basis of the estimated
value of
$ \emph N$.
In event-driven wireless sensor network, network congestion becomes severe when
an event is detected by number of sensors close to it. This subset of nodes
suddenly get activated to report the event via multi-hop transmission to a
distant
sink node or base station node. This type of traffic characteristics leads to
synchronized transmissions from the active sensors at the same time. Thus
population size of contending sensors that are reporting an event at a time
increases significantly. Hence, collision ratio is higher on detection of an
event in the environment. 

The BEB algorithm used in IEEE 802.11 DCF is not suitable for event-driven WSN.
First, Since BEB algorithm doubles its CW size after each collision, when active
population of nodes is large, it takes time to adapt to right value of CW size
after significant amount of collisions. It will happen when several nodes
observe an event around it the same time. Second, contention
window size is reset to its initial value $(W_{min})$ after each successful
transmission. This does not consider current status backlogged nodes indicated
by recently used CW. Larger CW is indication of large number of contending nodes
which may not settle down immediately in case of event-driven applications.
Finally, at the time of event occurrence, if contention window of the node is
already large, and only some sensor nodes get chance to report that event, then
delivery latency of the event is fairly high due to large CW size.
\section{Related Work}

In WSN, the research on MAC protocol design has been focused mainly on
energy-latency trade-offs. SMAC~\cite{ref10} is designed to save the energy by
using listen and sleep periodically with collision avoidance facilities of IEEE
802.11 standard. S-MAC uses synchronization mechanism to form virtual clusters
of sleep/wakeup schedule to avoid overhearing problem. Many variants of S-MAC
have been proposed to further decrease the energy consumption. These are
D-MAC~\cite{ref12}, T-MAC~\cite{ref13}, DS-MAC~\cite{ref13-1},
P-MAC~\cite{ref13-2}, TEEM~\cite{ref13-3}, DW-MAC~\cite{ref13-4} etc. These all
variants deal with major source of energy wastage such as idle-listening,
overhearing and collisions problems. B-MAC~\cite{ref13-5} is low power listening
mechanism , best suitable for low data rate wireless sensor networks. Similar to
S-MAC, and T-MAC, B-MAC also uses periodic ON and OFF of radio transceiver.
However, it is unsynchronized duty cycle protocol. In order to transmit the
packets, nodes sent long preamble before actual data transmission. These MAC
protocols have been designed for general sensor network applications where
latency is not considered as critical parameter.

In this paper, we are considering contention-based protocol for event-driven WSN
where latency is an important parameter. Many existing MAC protocols (e.g.
S-MAC, T-MAC, P-MAC etc.) utilize the IEEE 802.11 DCF mechanism in order to
handle hidden terminal, exposed terminal, and network congestion problems. As
discussed in previous section, performance of IEEE 802.11 DCF is weak when
network traffic is frequent or correlated depending on nodes' interaction with
physical environment. As a result, protocols based on IEEE 802.11 are not
suitable for event-driven WSN. In IEEE 802.11, backoff time is chosen randomly
in the range of [0, CW). The node waits for the chosen random number of vacant
slots before transmitting. If two nodes contend to access the channel at the
same time, and both find free channel for DIFS time, both of them transmit
resulting in collision. After collision, each colliding node choose random
number of vacant slots to wait from the range $[0, CW)$. Here CW for a node
is
$2^{i} \times W_{min}$, where $i_{th}$ collisions has been suffered for a
packet transmission. The node which has backoff time corresponding to lower slot
number gets the chance to access the channel while other nodes with higher
slot number wait for channel access. Backoff time is random
variable and can be a large value
depending on range from which random number is selected. For event-driven sensor
network applications, sometimes less number of nodes becomes active to report a
common event, or sometimes larger
number of nodes will report an event. So according to active population of nodes
reporting an event, a backoff algorithm must set right value of contention
window size.

For event-driven sensor network applications, very little work has been
published in literature. The CC-MAC was proposed by Vuran and
Akyldiz~\cite{ref14} for event-driven WSN. They have explored spatial
correlation in wireless sensor networks. In CC-MAC, Iterative Node
Selection (INS) algorithm was proposed to calculate correlation radius
($r_{corr}$)
based on correlation model. Only one node is allowed to transmit event
information within a correlation radius. In this way, CC-MAC suppresses the
transmissions by other nodes within same correlation radius. This single node is
referred as representative node selected by CSMA mechanism during each
transmission within $r_{corr}$ radius. In first contention phase, all nodes
within $r_{corr}$ radius contend to access the channel like in any other
contention based
protocol. As a result, only node winning the contention is selected as
representative node while other nodes turn to sleep. In CC-MAC, there is no
control on selection of representative node to further saving the energy. It
is unpredictable that which node will win the contention. SIFT MAC~\cite{ref15}
is
another protocol designed for event-driven WSN. The objective of SIFT MAC
protocol is to minimize the latency when spatially-correlated contention
occurs in a monitored area. Jamieson et. al~\cite{ref15} had argued that only R
nodes out of N nodes that report to a common event are sufficient to be
successful to transmit the event information to the sink node. SIFT MAC uses
non-uniform geometric distribution to choose slot number for picking up a slot
for transmission within fixed-size contention window (32 slots).


\section{Details of Adaptive Random Backoff (ARB)}
Since each event triggers a large number of nodes for sensing and transmission,
it is prudent to have a initial window (say $ W_{est} $) being estimated based
on past experience. Thus chance of collision are reduced to a greater extent.
Further CW can be adapted when collisions happen. When a node's transmission
collides $i^{th}$ times CW can be increased to $2^{i}W + W_{est}$. This
provides for desired adaptability. Here $W$ is increment in $CW$ after first
collision. The CW when packet is successfully transmitted can also be used to
update $W_{est}$ for use in future transmissions. In this section, we present
ARB scheme that is based on minor modifications of BEB. Unlike BEB, our ARB
protocol focuses on adjusting the contention window to achieve
higher throughput and better channel access delay for event-driven WSN
applications. 
 \subsection{Assumptions \& model used}
In the event-driven WSNs, we assume that network consisting of N sensor nodes,
is deployed with a sink node one hop away from sensor nodes. Based on Bianchi's
model~\cite{ref2} and ZA's model~\cite{ref5}, we analyze the performance
analysis of our modified protocol in order to optimize the channel access
specially for event-driven wireless sensor networks. We are interested in
specific area inside sensor field where an event is generated by means of some
activities for example detection of an intrusion or sudden change of temperature
in case of fire etc. When an event occurs significant number of sensor
nodes closer to it get activated and each triggered node has a packet to
transmit to sink node independently using MAC.

Following~\cite{ref2,ref7}, for a given node, let $w(t)$ the stochastic process
that represents $i^{th}$ backoff stage in the range $\emph i = 0,1,2, ...,i=L 
$, where L is maximum retransmission limit (retry limit) and $b(t)$ be the
sotchastic process that represents $k^{th}$ backoff counter in the range $\emph
k = 0,1,2, ...,W_{i}-1$,
where $W_{i}$ is given by$\left\lceil 2^{i}\times W_{min}\right\rceil$.
Therefore,
the system can be modeled as a Markov chain model $\{w(t),b(t)\}$ representing
the state of each node $\{i,k\}$. This model has assumption that each packet
collides with constant and independent probability $\emph p$ during each
transmission attempt irrespective of number of retransmission suffered, it has
been shown in~\cite{ref2,ref7} that: 
\subsubsection{\emph{the probability $\emph p$ can be expressed as }} 
\begin{equation}
\label{eq:thuput}
	\emph {p} = 1-(1-\tau)^{N-1}
\end{equation}
\subsubsection{\emph{the probability $\tau$ can be expressed as}} 
\begin{equation}
\label{eq:bkEq6}
	\tau = \sum_{i=0}^{m}b_{i,0}=b_{0,0}\frac{1-\emph{p}^{L+1}}{1-\emph{p}}
\end{equation}
where $b_{0,0} $ is given by Eq.~\eqref{eq:bkEq5}.
\begin{equation}
\label{eq:bkEq5}
b_{0,0} =
\frac{1}{\sum_{i=0}^{L}\left[1+\frac{1}{1-\emph{p}}\sum_{k=1}^{W_{i}-1}\frac{W_{
i}-k}{W_{i}} \right]\emph{p}^{i}}
\end{equation}

According to Bianchi's model~\cite{ref2}, the $\tau$ is known as attempt
probability.
Bianchi has used a two-dimensional model to obtain the expression with no
retransmission limit. Later, the limit for maximum retransmission was addressed
in ~\cite{ref7}. The Eq.~\eqref{eq:thuput} and Eq.~\eqref{eq:bkEq6} form a
system of two non-linear equations that has unique solution and can be solved
numerically for values of $\emph{p}$ and $\tau$.


\subsection{Proposed ARB algorithm for Event-Driven Scenario }
From analysis in~\cite{ref2}, BEB algorithm is main key
factor that influences the system efficiency. In IEEE 802.11 Binary Exponential
Backoff (BEB) algorithm, size of contention window (CW) is doubled on each
unsuccessful transmission and reset to initial value of $CW_{min}$ after each
successful transmission, described as Eq.~\eqref{eq:bkoff}. As illustrated
in \textbf{Algorithm 1}, the Backoff value $cw$ is randomly selected from
 the range $[0,CW]$, where $CW_{min} = 2^{i_{min}} - 1 $ and $cw = $ rand() mod
 $CW_{min}$, ($i_{min} = 5$ in 802.11 DCF). The CW will be double after each
unsuccessful
transmission attempt and continue to increase until it reaches the upper bound
$CW_{max}$, where $CW_{max} = 2^{i_{max}} - 1 $ (1023 in 802.11 DCF). When
retransmission limit reaches a threshold value, denoted as $m$, the CW doest not
increase further. If $i$ denotes the number of successive failed transmission
(due to collision or packet error), the increment in CW explained above can be
summarized:
\begin{equation}
\label{eq:bkoff}
		  W_{i} = \left\{ \begin{array}{ll}
         2^{i}\times W_{min} & \mbox{ $i \leq m$};\\
        2^{m}\times W_{min} & \mbox{ $i > m$}.\end{array} \right. 
\end{equation}
In the event-driven WSNs, contention resolution becomes critical due to
simultaneous transmission. To mitigate the severe collision into
sensing region, it is prudent to have average value of CW (say $ W_{avg}
$) being estimated based on past experience. The CW when packet is successfully
transmitted, can also be used to update $W_{avg}$ for use in future
transmissions. Therefore, the key idea of our ARB method is to
predict the $
W_{avg}$ based on current CW after each unsuccessful transmission attempt. 
Since it is based on minor modifications of BEB
algorithm, the backoff value is selected randomly from the contention
window. Unlike BEB algorithm, the lower bound of CW for next transmission is
updated with $CW_{avg}$ based on current contention window.
\subsubsection{Illustration of ARB}
\begin{figure}
 \centering

 \includegraphics[scale=0.60]{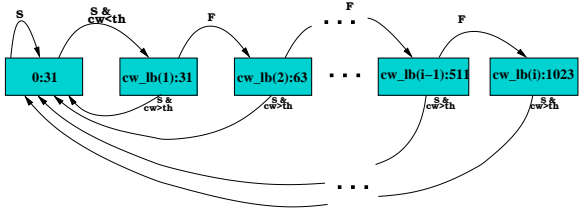}
  \caption{Proposed ARB algorithm for Event-Driven Scenario}
  \label{fig:BEB1}
 \end{figure}

ARB operates as follows (as shown in Fig.\ref{fig:BEB1} and \textbf{Algorithm
2}): when a node has a data packet to transmit a $cw_{i}$ is selected randomly
from $[0,CW_{min}]$ similar to BEB. Upon a successful data transmission, if
$cw_{i}$ is less
then $CW_{th}$ \footnote{$CW_{th}$ is used to
reset the lower bound of CW.}, a lower bound of CW for the next $cw_{i+1}$
selection will assigned as $CW_{lb} = \alpha \times CW_{avg}^{i}$, where
$CW_{avg}^i = 2 \times cw_{i} + CW_{avg}^{i-1}$. In case, the $cw_{i}$ is equal
to $0$, $CW_{lb}$ is set to default value \footnote{ Note that the default value
of $CW_{lb}$ is set to 1.}. Otherwise, $CW_{lb}$ will be reset to zero.
Therefore, the node will set the value of $cw_{i+1}$ from the range
$[CW_{lb},CW_{min}]$ for next transmission. If there is a failed
transmission, the CW is doubled and the backoff value, $cw$ is selected from
the range $[CW_{lb} - 1,
min(2^{i_{min}+n_f},2^{i_{max}}) - 1]$, where $n_f$ is the number of a failed
transmission. The CW keeps increasing until reaches the $CW_{max}$. 

\begin{algorithm}
\caption{Binary Exponential Backoff}
 \algsetup{indent=2em}
\begin{algorithmic}[1]
\STATE  $cw_{0} \leftarrow [0,2^{i_{min}} - 1]$

\FOR {each sending packet $P_{i}$} 
        \IF {fail in transmitting $P_{i}$}
 \STATE               $CW_{i+1} = 2 \times CW_{i}$
	\ELSE 
\STATE		$CW_{i+1} = 2 \times CW_{min}$
        \ENDIF
         \IF{$CW_{i+1} > CW_{max}$}
\STATE    $CW_{i+1} = CW_{max}$
        \ENDIF
 \STATE $cw_{i+1} \leftarrow [0, CW_{i+1} - 1]$
\ENDFOR 
\end{algorithmic}
\end{algorithm}

\begin{algorithm}
\caption{Adaptive Random Backoff For Event-Driven Scenario.}
 \algsetup{indent=2em}
\begin{algorithmic}[1]
\STATE  $ CW_{avg}^0=2 $
\STATE  $cw_{0} \leftarrow [0,2^{i_{min}} - 1]$
\FOR {each sending packet $P_{i+1}$} 
     \IF {$cw_i < CW_{th}$}
        \IF {$cw_i == 0$}
 \STATE               $CW_{avg}^i = 0$
	\ELSE 
\STATE		$CW_{avg}^i = 2 \times cw_{i} + CW_{avg}^{i-1}$
        \ENDIF
    \ELSE
\STATE               $CW_{avg}^i = 0$
  \ENDIF
 \STATE   $CW_{lb}^{i+1} = \alpha \times CW_{avg}^{i}$

 \STATE $cw_{i+1} \leftarrow [CW_{lb}^{i+1} - 1,
min(2^{i_{min}+n_f},2^{i_{max}}) - 1]$

\ENDFOR 
\end{algorithmic}
\end{algorithm}

In this way, the weighted average value of CW is computed dynamically in each
unsuccessful transmission attempt and kept unchanged upon successful
transmission attempt to use for future transmission. In addition, the collision
ratio is higher on detection of an event (explained in previous section).
After each successful transmission, BEB algorithm does not consider current
status of backlogged nodes indicated by current larger value of CW, it
decrements the CW to $CW_{min}$ immediately. To reduce the higher collision
ratio in the presence of an event, the $CW$ should follow gradually decrease by
halving the value on each successful transmission (similar to that
in~\cite{ref13-a,ref13-b}), instead of immediately resetting it to $CW_{min}$. 
%
 \subsection{Performance Evaluation }

The Eq.~\eqref{eq:thuput}, Eq.~\eqref{eq:bkEq6} and  Eq.~\eqref{eq:bkEq5} can be
rewritten for BEB model and our ARB model as follows:
\begin{equation}
\label{eq:bkEq7}
\dfrac{2(1-\emph{p})}{b_{0,0}} =
\sum_{i=0}^{L}\left[2(1-\emph{p})\emph{p}^{i}+(W_{i}-1)\emph{p}^{i}\right]
\end{equation}
 Where, $ W_{i}$ is contention window at $i^{th}$ backoff stage. Then the
relationship
between transmission Probability $\tau$ and conditional collision probability
$\emph{p}$ can be expressed by: 
\begin{equation}
\label{eq:bkEq8}
	\tau =
\dfrac{2(1-\emph{p}^{L+1})}{\sum_{i=0}^{L}\left[2(1-\emph{p})\emph{p}^{i}+(W_{i}
-1)\emph{p}^{i}\right]}
\end{equation}
 and 
\begin{equation}
\label{eq:bkEq9}
	\emph {p} = 1-(1-\tau)^{N-1}
\end{equation}

The performance analysis of BEB model and our ARB model have been carried out by
solving the Eq.~\eqref{eq:bkEq8} and  Eq.~\eqref{eq:bkEq9} numerically.
\subsubsection{Normalized System Throughput}
Let $ P_{tr}$, $P_{s} $ denote the probability that at least one transmission is
holding a given slot time and the probability of successful transmission, given
the probability $ P_{tr}$ respectively. So we have 
\begin{equation}
\label{eq:bkEq10}
	P_{tr} = 1-(1-\tau)^{N}
\end{equation}
\begin{equation}
\label{eq:bkEq11}
	P_{s} = \frac{N\tau(1-\tau)^{N}}{1-(1-\tau)^{N}}
\end{equation}
Let $S $ be the normalized system throughput which is the average payload sized
packet transmitted in a slot over the average duration of a slot time, expressed
as Eq.~\eqref{eq:bkEq12}.

\begin{align}
	S &= \frac{E[\mbox{Payload transmission during a slot
time}]}{E[\mbox{Duration of slot time}]} \nonumber \\
	&=
\frac{P_{s}P_{tr}E[P]}{P_{s}P_{tr}T_{s}+P_{tr}(1-P_{s})T_{c}+(1-P_{tr})T_{id}}
\label{eq:bkEq12}
\end{align}
where we use same symbols as used in paper~\cite{ref2}, $E[P]$ is average packet
payload size, $T_{s}$ is average time needed to transmit a packet of size
$E[P]$, $T_{id}$ is duration of idle period in a time slot, $T_{c}$ is average
time spent in collisions.
Let $T_{H}$, $T_{E[P]}$, $DIFS$, $SIFS$, $T_{ACK}$ denote the time to transmit
the header (including physical header, MAC header), time to transmit a payload
of size $E[P]$, DIFS time, SIFS time, and the time to transmit an ACK,
respectively. For the basic access method, we have
\begin{subequations}
\begin{align}
T^{\emph{basic}}_{\emph{s}}& = DIFS+ T_{H}+T_{E[P]}+\delta+SIFS+T_{ACK}+\delta
\label{eq:bkEq13a} \\
T^{\emph{basic}}_{\emph{c}}& = DIFS+
T_{H}+T_{E[P]}+SIFS+T_{ACK}\label{eq:bkEq13b}
\end{align}
\end{subequations}
Let $T_{\emph{RTS}}$ , $T_{\emph{CTS}}$ denote the time to transmit an RTS
packet and CTS packet, respectively. For the RTS/CTS access method, we have 

\begin{subequations}
\begin{align}
T^{\emph{RTS/CTS}}_{\emph{s}}& = DIFS+ T_{RTS}+SIFS+\delta+T_{CTS}+SIFS
\nonumber \\
&+\delta+T_{H}+T_{E[P]}+SIFS+\delta+T_{ACK}+2\delta \label{eq:bkEq14a} \\
T^{\emph{RTS/CTS}}_{\emph{c}}& = DIFS+ T_{RTS}+SIFS+T_{CTS}\label{eq:bkEq14b}
\end{align}
\end{subequations}


\subsubsection{Frame Dropping Probability}
Let $P_{\emph{drop}}$ be the probability that a packet is dropped. According to
the model, if a collision occurs a packet is dropped only in state \{L,0\}. So
frame dropping probability $P_{\emph{drop}}$ is
described as Eq.~\eqref{eq:bkEq15}. 
\begin{equation}
\label{eq:bkEq15}
	P_{\emph{drop}} = \emph{p}^{L+1}
\end{equation}

\subsubsection{Expected Channel Access Delay}
 We define channel access delay as the total time elapsed between instant of
time the packet comes in queue and time the packet gets chance of successful
transmission. In our model, we obtain the channel access delay by analyzing
expected backoff delay, transmission delay, and inter frame spaces (SIFS).
The expected backoff delay depends on backoff counter value and duration of
which counter freezes. Let total backoff slots be a random variable represented
by X without considering backoff slots for which a node freezes counter. The
probability that a packet is transmitted successfully after $\emph i_{th}$
retransmission, is given by $\frac{\emph{p}^{i}(1-\emph{p})}{1-\emph{p}^{L+1}} $
and average backoff slots are $\sum_{i=0}^{L}\frac{W_{i}-1}{2}$. So we have

\begin{equation}
\label{eq:bkEq16}
	E[X] =
\sum_{i=0}^{L}\frac{\emph{p}^{i}(1-\emph{p})}{1-\emph{p}^{L+1}}(\frac{W_{i}-1}{2
})
\end{equation}

where $E[X]$ is expected backoff delay for only successful transmission. 
Let $B$ be random variable representing backoff slots when counter freezes. The
probability that $E[X]$ is decreased when channel is idle, is given by
$(1-\emph{p})$. So we have
\begin{equation}
\label{eq:bkEq17}
	E[B] = \dfrac{E[X]}{(1-\emph{p})}\emph{p}
\end{equation}

Let $E[L_{retry}]$ be expected delay for retransmission (retry). We have
\begin{equation}
\label{eq:bkEq18}
	E[L_{retry}]
=\sum_{i=0}^{L}\frac{i\emph{p}^{i}(1-\emph{p})}{1-\emph{p}^{L+1}}
\end{equation}
Let D be a random variable representing transmission delay or packet delay. The
$T_{0}$ is the time when a node has to wait after its packet collision, and
$T_{ACK\_timeout}$, $T_{CTS\_timeout}$ represent ACK and CTS timeout duration
respectively. So we have
\begin{align}
	E[D]
&=E[X]\delta+E[B][(\frac{P_{s}}{P_{tr}})T_{s}+(\frac{P_{tr}-P_{s}}{P_{tr}})T_{c}
] \nonumber \\
&+E[L_{retry}](T_{c}+T_{o})+T_{s} \label{eq:bkEq19}
\end{align}
where average slot lengths are $\delta$ for an idle slot, $
\{(\frac{P_{s}}{P_{tr}})T_{s}+(\frac{P_{tr}-P_{s}}{P_{tr}})T_{c}\} $ for busy
slot, $(T_{c}+T_{o}) $ for failed transmission slot and $T_{s} $ for successful
transmission slot. 
Since channel is sensed busy when at least one transmission is holding a given
time slot, the probability that channel is busy in a time slot is same as
probability of successful transmission $ P_{tr}$. 
\begin{equation}
\label{eq:bkEq21}
	T^{\emph{basic}}_{\emph{o}}=SIFS+T_{ACK\_timeout}
\end{equation}
\begin{equation}
\label{eq:bkEq20}
	T^{\emph{RTS/CTS}}_{\emph{o}}=SIFS+T_{CTS\_timeout}
\end{equation}
%

\begin{table}
\caption{System Parameters for MAC and DSSS PHYS layer}
\centering
\begin{tabular}{|@{}r|@{}c|@{}r|@{}c|}
\hline 
SIFS     	&$10 \mu s$ &DIFS    	 &$50 \mu s$ \\  \hline
Slot time 	 & $20 \mu s$ & PHYS header     &$192 bits$ \\ \hline 
MAC header 	  &$224 bits$  &UDP/IP header     &$320 bits$ \\  \hline
ACK packet 	    &$112 bits$ &Data rate   	  &$11 Mbps$ \\ \hline 
Propagation delay     &$1 \mu s$ &Control rate   	 & $1 Mbps$  \\  \hline
Retry limit   	  &$6$  &CWmin, CWmax     &$32, 1024$  \\ \hline
\end {tabular}
\end{table}
\begin{table}[h!t]
\caption{Simulation and analytical results (LEGEND:NST,Normalized System
throughput(bps);CAD,Channel Access delays(second);N,Number of active reporting
nodes;S,Simulation;A,analytical;E,Relative Error)}
\centering
\begin{tabular}{ @{}c|@{}c|@{}r|@{}r|@{}r|@{}r|@{}r }
\hline 
$NST$ &	$S/A$	 & $N=50$	& $N=100$ 	& $N=200$	   & $N=300$
& $N=400$	 \bigstrut \\ \hline
     &	 $S$  & 701360	 & 515964	   & 430202	   & 368480	& 283290
\\  \cline{2-7}
     &	 $A$  & 650260	 & 579680	   & 502340	& 454700	   &
420300 \\  \cline{2-7}
     &	 $E$  & $0.07\% $	 & $-10.9\% $ 	   & $-14.3\% $	& $-18.9\% $   
& $-25.4\% $ \\ \hline
$CAD$ &	 $S$  & 7.349	 & 10.07	   & 11.53	& 18.33	   & 24.28\\
\cline{2-7}
     &	 $A$  & 6.175	 & 9.11	   & 13.56	& 17.24	   & 20.56\\ \cline{2-7}
     &	 $E$  & $19\% $	 & $10.5\% $	   & $-14.9\% $	& $6.32\% $	   &
$18\%$\\ \hline
\end {tabular}
\end{table}
We have evaluate the performance of ARB by comparing with existing BEB algorithm
through following parameters: initial window size $(T_{CW}) $, backoff
increasing factor $(\sigma)$, and retry limit $(L_{retry}) $ for basic access
method only due to space limitation. 
 

\section{Simulation \& Numerical Results}
In this section, we present simulation and Numerical results to evaluate the
performance of the proposed ARB. The analytical results were produced by solving
non-linear equations of model using MATLAB, then simulations were conducted
using ns-2 network simulator~\cite{ref16} to validate the analytical results.
All the parameters used in analytical model and our simulations can be found
in~\cite{ref2} for DSSS as summarized in Table-I. This paper uses simulation
model with similar assumptions as those in analytical model. Lucent's WaveLAN
parameters are used for radio model with 250 transmission range and 1 Mb/s
channel capacity. 

One hundred sensor nodes are randomly located in as area of 1000 m by 1000 m.
each node has enough data to send to obtain saturated condition. The number of
active reporting nodes that are engaged to report data for an event, can vary to
see the performance degradation due to increased collision probability. Using
ns-2, The IEEE802.11 simulation module, the DSDV routing module, and UDP
transport module had been used to configure the connections among sensor nodes.
For proposed ARB algorithm, the DSDV routing protocol and UDP transport module
had been used to create traffic connections among sensor nodes. The traffic
sources generate constant bit rate (CBR) traffic with the packet size of 1000
bytes at a rate of 2 packets/sec. for simulation time of 300 second. Table-II
shows the simulation versus numerical results for proposed ARB algorithm, where
the relative error is calculated by $\emph{(Simulation Result - Analytical
Result)/Analytical Result}$. As shown in the table, both the results are much
satisfactory with minimum errors for normalized system throughput (NST) and
channel access delays (CAD). 
\begin{figure}[t]
\centering
  \includegraphics[trim=1cm 0.5cm 1.25cm 0.45cm, clip=false,
scale=0.40]{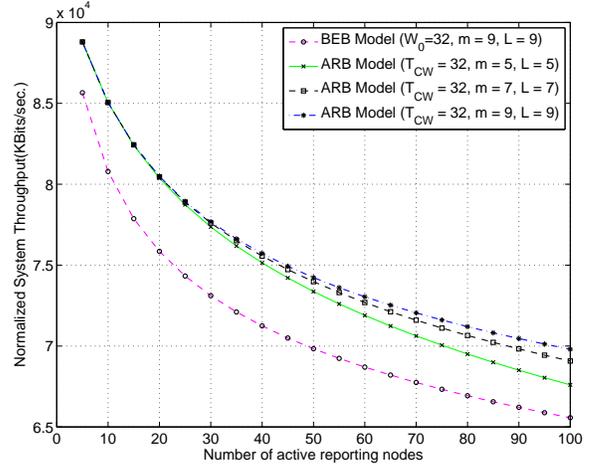}
  \caption{Normalized System Throughput vs. the number of active reporting nodes}
  \label{fig:thuputplot}
 \end{figure}
 \begin{figure}[h]
 \centering
 \includegraphics[trim=0.7cm 0.5cm 1.25cm
0.5cm,clip=false,scale=0.40]{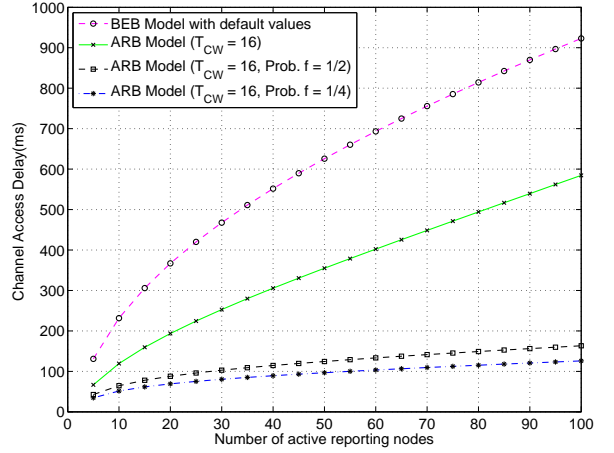}
  \caption{Channel Access Delay vs. the number of active reporting nodes}
  \label{fig:delayplot}
 \end{figure}
\begin{figure}[h]
 \centering
 \includegraphics[trim=0.7cm 0.7cm 1.25cm 0.45cm,
clip=false,scale=0.40]{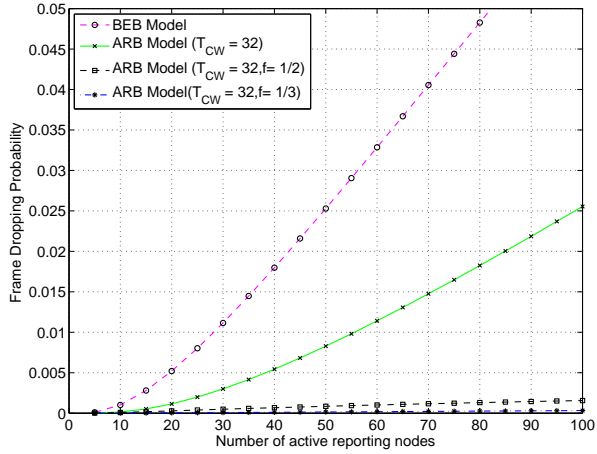}
  \caption{Frame Dropping Probability vs. the number of active reporting nodes}
  \label{fig:Frameplot}
\end{figure}
\begin{figure}[h!t]
\centering
  \includegraphics[trim=1cm 0.5cm 1.25cm 0.45cm, clip=false,
scale=0.40]{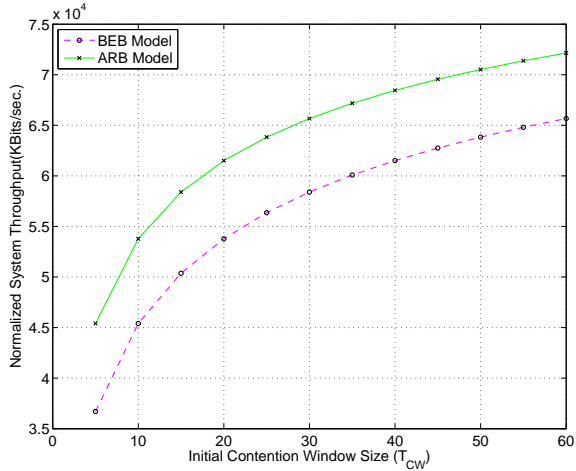}
  \caption{Normalized System Throughput vs. the initial contention window size
$(T_{CW})$}
  \label{fig:thuput1plot}
 \end{figure}
\subsection{Effectiveness of proposed ARB algorithm :}
Figs.\ref{fig:thuputplot}-\ref{fig:Frameplot} have results with following
parameters for BEB model and proposed ARB model:
$[\sigma_{\emph{BEB}},\sigma_{\emph{ARB}}]=[2,2]$,
$[L_{retry,\emph{BEB}},L_{retry,\emph{ARB}}]=[4,4]
$,$[W_{0,\emph{BEB}},T_{CW}]=[32,16]$, and
$[CW_{max,\emph{BEB}},CW_{max,\emph{ARB}}]=[1024,1024]$.
Figs.\ref{fig:thuputplot}-\ref{fig:Frameplot} show normalized system throughput,
Channel access delay and Frame-dropping probability for BEB model and proposed
ARB model over the number of active reporting nodes. As shown in Figures, proposed
ARB model has better throughput (channel access delay) than that of BEB model of
IEEE 802.11 DCF. It is clearly shown that proposed ARB model with initial
contention window $(T_{CW})$ and halving contention window with probability
$\textbf{\emph{f}}$ after each successful transmission, contributes to further
improvement in the performance. 
 \begin{figure}[t]
 \centering
 \includegraphics[trim=0.79cm 0.69cm 1.25cm 0.45cm,
clip=false,scale=0.40]{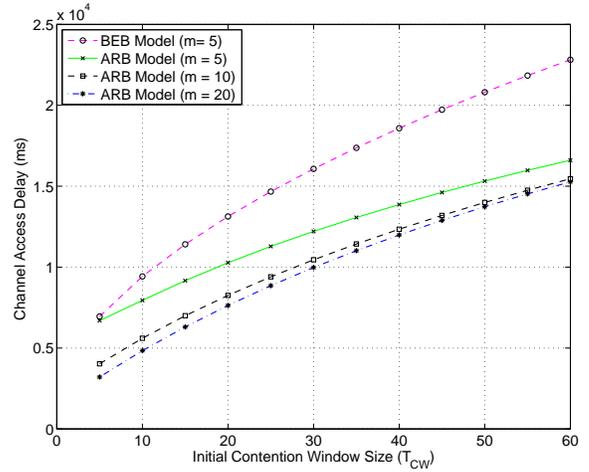}
  \caption{Channel Access Delay vs. the initial contention window size
$(T_{CW})$ }
  \label{fig:delay1plot}
 \end{figure}
\begin{figure}[h]
 \centering
 \includegraphics[trim=0.79cm 0.69cm 1.25cm 0.45cm,
clip=false,scale=0.40]{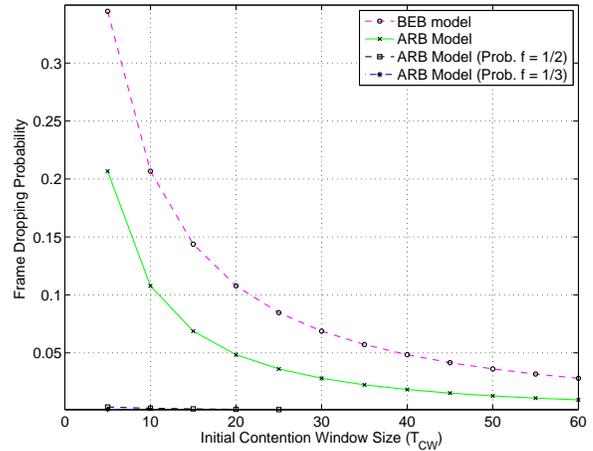}
  \caption{Frame Dropping Probability vs. the initial contention window size
$(T_{CW})$}
  \label{fig:Frame1plot}
\end{figure}
 \begin{figure}[h!t]
 \centering
 \includegraphics[trim=0.79cm 0.69cm 1.25cm 0.45cm, clip=false,
scale=0.40]{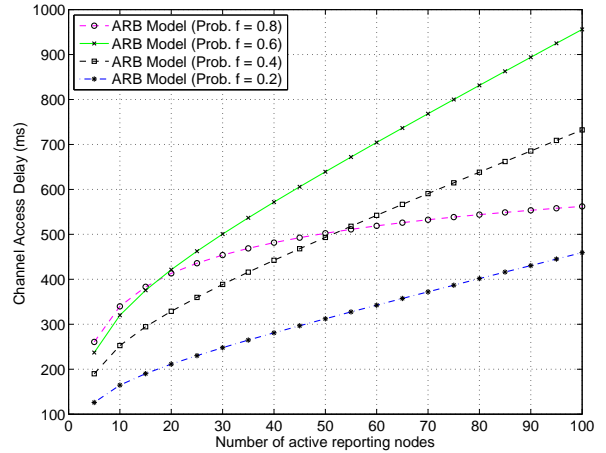}
  \caption{Channel Access Delay with different probabilities $\textbf{\emph{f}}$
over the number of active reporting nodes}
  \label{fig:delay2plot}
 \end{figure}
\subsection{Effects of initial window size $(T_{CW})$ in ARB algorithm:}
Figs.\ref{fig:thuput1plot}-\ref{fig:Frame1plot} have results with following
parameters for proposed ARB model: $\sigma_{\emph{BEB}}=2,
L_{retry,\emph{ARB}}=4, CW_{max,\emph{ARB}}=1024,$ and $N=100$.
Fig.\ref{fig:thuput1plot} and Fig.\ref{fig:delay1plot} show normalized system
throughput, Channel access delays over the initial contention window size
$(T_{CW})$that changes from 4 to 60. As shown in Fig.\ref{fig:thuput1plot} and
Fig.\ref{fig:delay1plot}, as the initial contention window $(T_{CW})$ increases
the throughput (channel access delay) increases, but larger value of $T_{CW}$
will increase the number of free slots which causes the channel access delay.
The results using proposed ARB model indicate that better throughput (channel
access delay) can be achieved with minimum value of initial contention window
size $(T_{CW})$. Fig.\ref{fig:Frame1plot} shows that as the initial CW size 
increases frame-dropping probability decrease. The frame-dropping probability
can be minimized by selecting an optimal value of probability
$\emph{f}$ between $\emph{Prob. f = 1}$ and
$\emph{Prob. f = 0.5}$. 

\subsection{Effects of halving the CW with probability $\textbf{f}$ in ARB
algorithm:}
Figs.\ref{fig:delay2plot}-\ref{fig:Frame2plot} have results with following
parameters for our ARB model: $\sigma_{\emph{BEB}}=2, T_{CW}=4,
L_{retry,\emph{ARB}}=4,$ and $CW_{max,\emph{ARB}}=1024$.
Fig.\ref{fig:delay2plot} shows Channel access delays with different probabilities
$\textbf{\emph{f}}$ over the number of active reporting nodes. After each
successful transmission, halving the CW will minimize the new collisions so that
proposed ARB model using this enhancement will decrease the channel access delay as the number of active reporting nodes increase.

\section{Conclusions}
In the design of a backoff algorithm for IEEE 802.11 DCF based event-driven WSN,
the contention window size must be adjusted according to the number of reporting
nodes of an event which is indication of current network status. So that network
performance can be improved to report data of an event into sensor
network applications. We have presented and analyzed two enhancements of 802.11
MAC in order to optimize the channel access over event-driven workload. Using Bianchi's analytical model,
we analyze the performance of existing BEB and proposed ARB algorithm to obtain
network throughput, channel access delay and frame-dropping probability. To
validate the analytical results, the simulation results are provided. The
results indicate that proposed ARB algorithm yielded higher network throughput
and lower channel access delay than backoff algorithm (BEB) of IEEE 802.11 MAC. These changes could reduce the average
number of free slots and collision probability. Our results show that with halving CW according to probability $\textbf{f}$,
both throughput and latency can be improved with a higher frame-dropping
probability. 





%

\end{document}